\documentclass[twocolumn,showpacs,preprintnumbers,amsmath,amssymb,prl,superscriptaddress]{revtex4}

\usepackage{graphicx}

\begin{document}

\title{Dual fermion approach to nonlocal correlations in the Hubbard model}

\author{A. N. Rubtsov}
\affiliation{Department of Physics, Moscow State University,
119992 Moscow, Russia }
\author{M. I. Katsnelson}
\affiliation{Institute for Molecules and Materials, Radboud
University, 6525 ED Nijmegen, The Netherlands}
\author{A. I. Lichtenstein}
\affiliation{Institute of Theoretical Physics, University of
Hamburg, 20355 Hamburg, Germany} 

\begin{abstract}
A new diagrammatic technique is developed to describe nonlocal
effects (e.g., pseudogap formation) in the Hubbard-like models. In
contrast to cluster approaches, this method utilizes an exact
transition to the dual set of variables, and it therefore becomes
possible to treat the irreducible vertices of an effective {\it
single-impurity} problem as small parameters. This provides a very
efficient interpolation between weak-coupling (band) and atomic
limits. The antiferromagnetic pseudogap formation in the Hubbard
model is correctly reproduced by just the lowest-order diagrams.
\end{abstract}


\maketitle

Lattice fermion models with a strong local interaction
(Hubbard-like models \cite{Hubbard63}) are believed to catch the
basic physics of various systems, such as high-temperature
superconductors \cite{Scalapino,Anderson}, itinerant-electron
magnets \cite{Moriya}, Mott insulators \cite{Georges}, ultracold
atoms in optical lattices \cite{ColdGases}, etc. Unfortunately,
the analytical treatment of these problems is essentially
restricted by the lack of explicit small parameters for the most
physically interesting interactions. Direct numerical methods,
such as exact diagonalization \cite{exactdiag} or quantum Monte
Carlo (QMC) \cite{HirshFye,ScalapinoQMC} are limited by the
clusters being of a relatively small size, or face other obstacles
such as the famous sign problem for QMC simulations at low
temperature \cite{deReadt}. There is a very successful approximate
way to treat these models via the framework of so-called dynamical
mean-field theory (DMFT) \cite{Georges}, where the lattice
many-body problem is replaced with an effective impurity model.
This approach is essentially based on the assumption of a local
(i.e. momentum-independent) fermionic self energy. Indeed, there
are numerous interesting phenomena which are basically determined
by {\it local} electron correlations, such as Kondo effect
\cite{Hewson}, Mott-Hubbard transitions \cite{Georges} and local
moment formation in itinerant-electron magnets
\cite{LichtPRL2001}. At the same time, momentum dependence of the
self energy is of crucial importance for Luttinger liquid
formation in low-dimensional systems \cite{Anderson,Mahan}, d-wave
pairing in high-$T_c$ superconductors
\cite{Scalapino,JarellRMP,Licht2000}, and non-Fermi-liquid
behavior due to van Hove singularities in two dimensions
\cite{Katanin}. Recently a rather strong momentum dependence of
the effective mass renormalization in photoemission spectra of
iron was observed \cite{FeARPES}.

Currently, non-local many body effects in strongly correlated
systems are mainly studied via the framework of various cluster
generalizations of DMFT
\cite{JarellRMP,Licht2000,KotliarCell,Potthoff}. Cluster methods
do catch basic physics of d-wave pairing and antiferromagnetism in
high-$T_c$ superconductors \cite{JarellRMP,Licht2000}, and the
effects of intercite Coulomb interaction in various
transition-metal oxides \cite{Mazurenko,PoterTi2O3,PoterVO2}. At
the same time, however effects like Luttinger liquid formation or
van Hove singularities can not be described in cluster approaches.
In such cases the correlations are essentially long-ranged and it
is more natural to describe the correlations in momentum space.
Recently attempts have been made to consider non local correlation
effects in momentum space starting from DMFT as a zeroth-order
approximation \cite{Held,Japan}. This approach requires a solution
of ladder-like integral equation for complete vertex $\Gamma$ and
the subsequent use of the Bethe-Salpeter equation to obtain
Green's functions. The first step here exploits an irreducible
vertex of the effective impurity problem $\gamma^{(4)}$, whereas
the second step uses just the bare interaction parameter $U$. This
second step makes the generalized DMFT approoach mostly suitable
to the weak-coupling regime \cite{explain}.

In this Letter, we present a scheme which is accurate in both
small-$U$ and large-$U$ limits and does not require numerically
expensive solutions of any integral equations. A comparison of the
results with lattice QMC simulations for the two-dimensional (2D)
Hubbard model in the pseudogap regime demonstrates that the scheme
is actually accurate even in the less-favorable case of
intermediate $U$.

We proceed with 2D Hubbard model with the corresponding
imaginary-time action
\begin{equation}\label{Hubbard}
    S[c,c^*]=\sum_{\omega k \sigma}
    \left(\epsilon_k-\mu-i \omega\right) c^*_{\omega k \sigma} c_{\omega k \sigma} + U \sum_i \int_0^\beta n_{i\uparrow\tau} n_{i\downarrow\tau} d\tau.
\end{equation}
Here $\beta$ and $\mu$ are the inverse temperature and chemical
potential, respectively, $\omega=(2 j+1) \pi/\beta, j=0,\pm 1,
...$ are the Matsubara frequencies, $\sigma=\uparrow, \downarrow$
is the spin projection. The bare dispersion law is $\epsilon_k=-2
t (\cos k_x+\cos k_y)$, $c^*, c$ are the Grassmannian variables,
$n_{i \sigma \tau}=c^*_{i \sigma \tau} c_{i \sigma \tau}$, where
the indices $i$ and $k$ label sites and quasi-momenta.

In the spirit of DMFT, we introduce a single-site reference system
(an effective impurity model) with the action
\begin{equation}\label{Impurity}
S_{imp}=\sum_{\omega,\sigma} (\Delta_ \omega-\mu-i \omega)
c^*_{\omega,\sigma} c_{\omega,\sigma} + U\int_0^\beta n_{\uparrow
\tau} n_{\downarrow \tau}  d\tau
\end{equation}
where $\Delta_\omega$ is as an yet undefined hybridization
function describing the interaction of the effective impurity with
a bath. We suppose that all properties of the impurity problem are
known, so that its single-particle Green's function $g_w$ is
known, and the irreducible vertex parts $\gamma^{(4)},
\gamma^{6},$ etc. Our goal is to express the Green's function
$G_{\omega k}$ and vertices $\Gamma$ of the lattice problem in
Eq.(\ref{Hubbard}) via these quantities.

Since $\Delta$ is independent of $k$, the action (\ref{Hubbard})
can be represented in the form
\begin{equation}\label{Hubbard2}
    S[c,c^*]=\sum_i S_{imp}[c_i, c_i^*]-\sum_{\omega  k \sigma}
    (\Delta_\omega-\epsilon_k) c^*_{\omega k \sigma} c_{\omega k
    \sigma}.
\end{equation}

\begin{widetext}
We utilize a dual transformation to the set of new Grassmannian
variables $f, f^*$. The following identity
\begin{equation}\label{Gauss}
    e^{ A^2 c^*_{\omega k \sigma} c_{\omega k \sigma} }= B^{-2}
\int e^{- A B (c^*_{\omega k \sigma} f_{\omega k \sigma} +
f^*_{\omega k \sigma} c_{\omega k \sigma}) - B^2  f^*_{\omega k
\sigma} f_{\omega k \sigma}}  d f^*_{\omega k \sigma} d f_{\omega
k \sigma},
\end{equation}
is valid for arbitrary complex numbers $A$ and $B$. We chose
$A^2=(\Delta_\omega-\epsilon_k)$ and $B^2= g^{-2}_\omega
(\Delta_\omega-\epsilon_k)^{-1}$ for each set of indices $\omega,
k, \sigma$.

With this identity, the partition function of the lattice problem
$Z=\int  e^{-S[c,c^*]} {\cal D} c^* {\cal D} c$ can be presented
in a form $Z=Z_f \int \int e^{-S[c,c^*,f,f^*]} {\cal D} f^* {\cal
D} f {\cal D} c^* {\cal D} c$, where

\begin{equation}\label{Scf}
    S[c,c^*,f,f^*]=\sum_i S_{imp}[c_i, c_i^*]+\sum_{\omega k \sigma}
     \left[g_\omega^{-1} (f^*_{\omega k\sigma} c_{\omega k\sigma} + c^*_{\omega k \sigma} f_{\omega k \sigma})
    +g_\omega^{-2} (\Delta_\omega-\epsilon_k)^{-1}
    f^*_{\omega k \sigma} f_{\omega k \sigma}\right]
\end{equation}
\end{widetext}
and $Z_f$ is a product $\Pi_{\omega k} g_\omega^{2}
(\Delta_\omega-\epsilon_k)$.

As a  next step, we establish an exact relation between the
Green's function of the initial system $G_{\tau-\tau', i-i'}=-< T
c_{\tau i} c^*_{\tau' i'} >$ and that of the dual system
$G^{dual}_{\tau-\tau', i-i'}=- <T f_{\tau i} f^*_{\tau' i'}>$. To
this aim, we can replace $\epsilon_k \to \epsilon_k+\delta
\epsilon _{\omega k}$ with a differentiation of the partition
function with respect to $\delta \epsilon _{\omega k}$. Since we
have two expressions for the action (\ref{Hubbard}) and
(\ref{Scf}), one obtains
\begin{equation}\label{exact}
    G_{\omega, k}=g_\omega^{-2} (\Delta_\omega-\epsilon_k)^{-2}
    G^{dual}_{\omega, k}+(\Delta_\omega-\epsilon_k)^{-1},
\end{equation}
where the last term follows from the differentiation of $Z_f$.

The crucial point is that the integration over the initial
variables $c^*_i, c_i$ can be performed separately for each
lattice site, since $\sum_k \left( f^*_k c_k + c^*_k f_k \right) =
\sum_i \left( f^*_i c_i + c^*_i f_i \right)$. For a given site
$i$, one should integrate out $c^*_i, c_i$ from the action that
equals
$S_{site}[c_i,c_i^*,f_i,f_i^*]=S_{imp}[c_i,c^*_i]+\sum_\omega
g^{-1}_\omega(f^*_\omega c_\omega + c^*_\omega f_\omega)$. We
obtain
\begin{equation}
\int e^{-S_{site}} {\cal D} c_i^* {\cal D} c_i=Z_{imp}
e^{-\sum_{\omega \sigma} g^{-1}_{\omega} f^*_{\omega i \sigma}
f_{\omega  i \sigma} - V[f_i, f^*_i]},
\end{equation}
where $Z_{imp}$ is a partition function of the impurity problem
(\ref{Impurity}). The Taylor series for $V[f_i, f^*_i]$ in powers
of $f_i, f^*_i$ starts from $-\gamma^{(4)}_{1234} f^*_1 f_2 f^*_3
f_4$ (indices stand for a combination of $\sigma$ and $\omega$,
for example $f_1^*$ means $f_{\sigma_1, \omega_1}^*$). Further
Taylor series terms yeld similar combinations including
$\gamma^{(n)}$ of higher orders.

We arrive with an action $S$ depending on the new variables $f,
f^*$ only:
\begin{equation}\label{Sf}
    S[f,f^*]= \sum_{\omega k \sigma}g_\omega^{-2} \left((\Delta_\omega - \epsilon_k)^{-1}
    +g_\omega\right) f^*_{\omega k \sigma} f_{\omega
    k \sigma} + \sum_i V_i,
\end{equation}
with $V_i\equiv V[f^*_i,f_i]$. In this dual action, the
interaction terms remain localized in space, but are non-local in
imaginary time, since, for example $\gamma^{(4)}$ depends on the
three independent Matsubara frequencies. To obtain the dual
potential $V$ for a practical calculation, one should solve then
the impurity problem (\ref{Impurity}).

\begin{figure}
\includegraphics[width=\columnwidth]{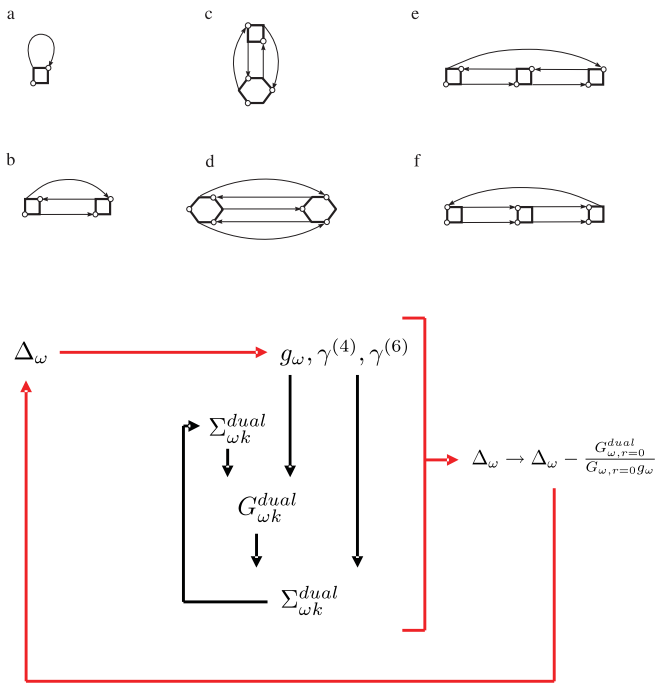}
\caption{(color online) Various diagrams for $\Sigma^{dual}$ and
the scheme of calculation. The calculation includes ``big'' and
``small'' loop, marked with red and black lines, respectively. The
small loop is to determine the renormalized dual Green's function
$G^{dual}$ in a self-consistent way, for given $\Delta, g$, and
$\gamma^{(n)}$. The big loop is to determine $\Delta$. Only the
big loop requires a solution of the impurity problem.}
\end{figure}

Finally a regular diagrammatic expansion in powers of $V$ can be
performed. We draw skeleton diagrams, so that the lines in
diagrams are renormalized dual Green's function, whereas the
vertices are $\gamma^{(n)}$. The rules of diagram construction are
very similar to usual ones, but the six-leg and higher-order
vertices appear because $\gamma^{(6)}$ and higher terms are
present in the series for $V$. Figure 1 shows several diagrams
contributing dual self-energy $\Sigma^{dual}_{\omega,
k}=-[(\Delta_\omega-\epsilon_k)^{-1}g^{-2}_\omega+g^{-1}_\omega+(G^{dual}_{\omega,
k})^{-1}$].

We use the skeleton-diagram expansion for the dual self-energy
since it leads to the conserving theories, exactly like in
conventional diagram technique \cite{Luttinger,Baym,Bickers}.
The Baym criterion of a conservative theory is the existence of a
functional of the Green function $\Phi[G]$ such that $\frac{\delta
\Phi}{\delta G}=\Sigma$. Here, the variation $\delta G$ comes from
the infinitesimal variation of the Gaussian part of the action,
$\delta (G^{0})^{-1} c^* c$. In our consideration, we  consider
also the infinitesimal variations of the dual potential $\delta
(G^{0}_{dual})^{-1} c^* c$. One can call an approximation dually
$\Phi$-derivable, if there exists a functional $\Phi^{dual}
[G^{dual}]$ such as $\frac{\delta \Phi^{dual}}{\delta
G^{dual}}=\Sigma^{dual}$, where the variation comes from $\delta
(G^{0}_{dual})^{-1}$. Now, it turns out that the theory is
$\Phi$-derivable if it is dually $\Phi$-derivable. The proof uses
the relation between functional $\Phi$ and the partition function
$\ln Z=\Phi-{\rm Tr} \Sigma G -{\rm Tr} \ln (-G)+C$ (here $C$ is
an additive constant; see Ref.\onlinecite{Baym}, Eq.(47)). Since a
similar relation takes place for $\Phi^{dual}$ and $\ln Z$ and
since the partition function is the same for the initial and dual
variables, this gives a one-to-one correspondence between
$\Phi[G]$ and $\Phi^{dual}[G^{dual}]$. This is just a sketch; the
detailed proof will be published elsewhere.


It is important to understand what can be a small parameter in the
expansion in dual diagrams. Clearly, if $U$ is small, then
$\gamma^{(4)}\propto U, ~ \gamma^{(6)}\propto U^2$ etc., and in
the weak-correlated regime vertices in the diagrams will be small
(Fig. 1), and higher-order vertices will be even smaller.

At this point we establish a condition for $\Delta$, which was so
far an arbitrary quantity. We use a self-consistent condition
\begin{equation}\label{NoLoops}
    \sum_k G^{dual}_{\omega, k}=0.
\end{equation}
It means that the simple closed loops in diagrams vanish. In
particular, this leads to the vanishing of the first-order
``Hartree'' corrections in the diagrammatic expansion. The diagram
series of this kind has several important peculiarities. First of
all, let us consider the zeroth-order approximation,
$\Sigma^{(dual)}=0$. In this case, the condition (\ref{NoLoops})
becomes
\begin{equation}\label{NoLoopsDMFT}
    \sum_k \frac{1}{g_\omega+(\Delta_\omega-\epsilon_k)^{-1}}=0.
\end{equation}
It is easy to show that this is exactly equivalent to the DMFT
equation for the ``hybridization function'' $\Delta_\omega$
\cite{Georges}. It is known that DMFT behaves correctly near the
atomic limit. In terms of the dual variables, one can observe that
since $\epsilon, \Delta \ll g^{-1}$ near the atomic limit, it
follows from the condition (\ref{NoLoopsDMFT}) that
$G^{dual}\approx g^2_\omega \epsilon_k<<g_\omega$ in this case. It
easy to check that this argumentation is valid the scheme of an
arbitrary diagrammatic order: the dispersion of $G^{dual}$ is
small near the atomic limit and therefore (\ref{NoLoops}) means
that lines in dual diagrams carry a small factor $\epsilon
g^{-1}$. This ensures the fast convergence of new diagrammatic
expansion in the strong-coupling limit.

As the most challenging test of our approximation scheme in the
intermediate regime, we performed the calculation for the
half-filled square-lattice Hubbard model, at sufficiently low
temperature $\beta^{-1}=|t|/5$. The value of U was varied from
small numbers to a bandwidth $8 |t|$. The block scheme of our
calculation is shown in Fig. 1. It has a good practical
convergence: typically, about 10 iteration are enough to ensure
convergence.

In order to obtain reference point for a further comparison with
the results of our new approximation scheme, we performed a direct
lattice QMC calculation with the continuous-time QMC code
\cite{ctqmc}. There are strong antiferromagnetic fluctuations in
the system, although true antiferromagnetism is impossible at
finite temperature in the 2D system with an isotropic order
parameter \cite{Mermin}. Consequently, the increase of $U$ results
in a formation of an antiferromagnetic pseudogap.

It was also noticed that single-site DMFT calculation for this
system shows no pseudogap in the density of states, although the
data for local part of self-energy are reproduced quite well in
DMFT. Thus we concluded that the formation of pseudogap is
entirely related to the non-local part of $\Sigma$, neglected in
DMFT.

We present the results of the dual-fermion calculation with only
diagram (b) taken into account. All other diagrams are smaller
both in the strong-coupling and weak-coupling regime, due to extra
vertices or extra lines, respectively.
Computational results are illustrated by Fig. 2. The upper panel
shows an imaginary part of the self-energy. In the DMFT this
quantity is momentum-independent. Our calculations show a very
strong ${\bf k}$-dependence with a maximum near the Fermi surface.
At relatively small value $U=1$ the peaks of ${\rm Im} \Sigma$ are
located near the van Hove singularities (left picture), as it can
be understood from the weak-coupling expansion. Contrary, for an
antiferromagnetic system near the atomic limit, ${\rm Im}
\Sigma_{k,\omega=0}$ would be a simple delta-function peaking at
Fermi surface. For a pseudogap regime at finite $\beta, U$, the
width of this peak is of course finite, but the altitude almost
does not depend on the point at Fermi surface (right picture). The
lower panel shows an effective renormalized dispersion law
$\epsilon_k+{\rm Re} \Sigma_{k,\omega=0}$. For the metallic
regime, the renormalization is small. For an antiferromagnetic
insulator, the would be a pole in ${\rm Re} \Sigma_{k,\omega=0}$
at the Fermi surface. For the pseudogap regime, fluctuations
virtually move this pole from the real-frequency axes, as the
curve for $U=2$ shows.

Thus, our scheme continuously interpolate between the two very
different regimes. It should be stressed that the quantities under
study have very strong $k$-dependence and that it would be very
difficult to obtain the result of this kind, for example, in
cluster calculations. Whereas for the weak-coupling regime
effective schemes to calculate nonlocal self-energy are known,
such as FLEX \cite{Bickers} or parquet \cite{Katanin}, to our
knowledge, there is no alternative scheme yet for the strong
coupling case.

\begin{figure}
\includegraphics[width=\columnwidth]{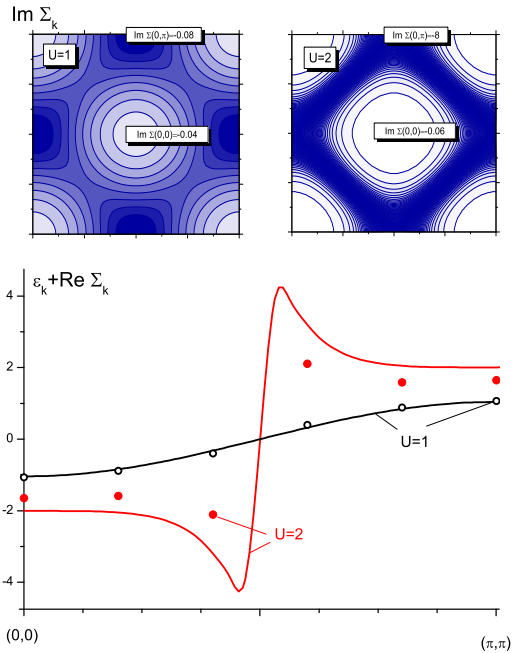}
\caption{(color online) Fermi-energy properties of the half-filled
Hubbard model calculated with the leading dual diagram correction
{\it b}. The calculations have been performed for the bandwidth $8
t=2$ at $\beta=20$, for different values of $U$. \textsc{Upper
panels} are contour plots for ${\rm Im} \Sigma_k$ at Fermi energy.
At $U=1$, ${\rm Im} \Sigma$ peaks in the four van Hove points,
whereas ${\rm Im} \Sigma (U=1)$ is approximately constant in all
points of the Fermi surface. Note also that the change from $U=1$
to $U=2$ leads to a $10^2$ increase in ${\rm Im} \Sigma$.
\textsc{Lower panel} shows a graph of the effective dispersion
law, $\epsilon_k+{\rm Im} \Sigma_k$ at Fermi level, plotted along
the $(0,0)~ - ~(\pi,\pi)$ direction. The initial ``cosine''
dispersion law $\epsilon_k$ is almost not renormalized at $U=1$.
Contrary, for $U=2$ the curve shows the antiferromagnetic
properties. The result of direct QMC $10\times 10$ lattice
simulation are shown with dots and confirm this picture.}
\end{figure}

To conclude, we have formulated an effective perturbation theory
to calculate the momentum dependence of self energy starting with
single-site DMFT or any local approximations. The vertices of the
effective impurity problem play the role of formal small
parameters. Due to the transformation to dual fermionic variables,
consideration of a few leading diagrams provides a quite
satisfactory description of the nonlocal correlation effects in a
broad range of parameters, up to the atomic limit. This scheme can
be easily generalized to multiband case to be implemented into
realistic electronic structure calculations for strongly
correlated systems.

The work was supported by NWO project 047.016.005 and FOM (the
Netherlands), DFG Grant  No. SFB 668-A3 (Germany), and Leading
scientific schools program and ``Dynasty'' foundation (Russia).


\begin{thebibliography}{99}
\bibitem{explain} Of course, the Bethe-Salpeter equation itself is exact
irrespective of the value of $U$, up to atomic limit. However, the
first step of the procedure produces an approximate vertex
$\Gamma$, which is insufficient in the case of large $U$.
\end{thebibliography}
\end{document}